\documentclass[aps,prd,onecolumn,amsmath,11pt,superscriptaddress,floatfix,nofootinbib,preprintnumbers]{revtex4-2}

\usepackage{mathrsfs}
\usepackage{amsfonts}
\usepackage{amsmath,amssymb,bm}
\usepackage{array}
\usepackage{verbatim}
\usepackage{epsfig}
\usepackage{graphicx}
\usepackage{hyperref}
\hypersetup{colorlinks, linkcolor = [rgb]{0, 0, 0.5}, citecolor = [rgb]{0,0.0,0.5}, urlcolor = [rgb]{0,0.0,0.5}}
\usepackage[normalem]{ulem}
\usepackage{xcolor}
\usepackage{ulem}
\usepackage{multirow}

\usepackage{subcaption}
\usepackage{graphicx}% Include figure files
\graphicspath{{./figs/}}
\usepackage{dcolumn}% Align table columns on decimal point
\usepackage{bm}% bold math
\usepackage{slashed}
\usepackage{siunitx}
\usepackage{ulem,xpatch}
\usepackage{hyperref}% add hypertext capabilities
\usepackage[mathlines]{lineno}% Enable numbering of text and display math
\usepackage{comment}
\usepackage{soul}
\usepackage{xcolor}
\usepackage{xfrac}

\allowdisplaybreaks[4]

\begin{document}

\title{Resolving Local Structure of Distribution Functions in Collider Measurements}

\author{Cong Li}
\affiliation{School of Information Engineering, Zhejiang Ocean University, Zhoushan, Zhejiang, China}
%\author{Jing Zhao}
%\affiliation{Key Laboratory of Particle Physics and Particle Irradiation (MOE), Institute of Frontier and Interdisciplinary Science, Shandong University, Qingdao, Shandong 266237, China}
%

%\date{\today}

\begin{abstract}
Collider measurements may blur narrow features in underlying distributions before fitting, since a data point samples a finite region. We show that observables linear in a selected target distribution, after externally constrained inputs are absorbed into a nonnegative finite response, have a calculable distribution-space locality set by hard dynamics and kinematics. The baseline-normalized $F_{\rm dir}(X)$ yields the local value at zero width and a controlled average otherwise, with covariance setting the leading bias. This guides the choice of processes and observables that retain unexpected fine structure. An EIC photon-radiation observable realizes the exact zero-width limit.

\end{abstract}

\maketitle
\section{Introduction}
\label{s.intro}
Collider experiments infer unobservable partonic, photonic, and other distributions from differential cross sections, making their extraction an inverse problem \cite{blobel2002unfoldingmethodhighenergy,annure}. We consider measurements that, after externally constrained inputs are absorbed into the theoretical response, can be written linearly in a selected target distribution as \cite{COLLINS1985104}
 \begin{equation} 
 \frac{d\sigma}{dX} = \int d\Phi\, H(X,\Phi)\, F\!\left[\xi(X,\Phi)\right]. 
 \label{eq:factorized_observable} 
 \end{equation} 
 Here, $X$ denotes the experimentally measured variables, $\Phi$ contains the unobserved or integrated-out degrees of freedom, $H$ is the calculable theoretical weight remaining after the unknown distribution $F$ has been separated, and $\xi$ denotes the distribution coordinate entering $F$. Because different unobserved configurations generally correspond to different values of $\xi$, a fixed measurement point $X$ does not in general probe a single point of $F$, but instead receives contributions from a finite region. 

From the perspective of information transfer and resolving power \cite{10.1111/j.1365-246X.1968.tb00216.x}, the hard-scattering process acts analogously to an imaging system: if one measurement point corresponds to a broad region of the distribution, local structures are smoothed when mapped onto the experimental measurement; if the corresponding region is sufficiently narrow, more local information can be directly preserved. This smoothing is distinct from finite statistics or detector resolution because it is already present in the theoretical forward mapping itself. Even for an ideal experimental measurement, integration over unobserved degrees of freedom still mixes information from different distribution coordinates. 

Global fits, unfolding, and other inversion methods \cite{blobel2002unfoldingmethodhighenergy} therefore remain essential tools \cite{DAGOSTINI1995487,PhysRevLett.124.182001,Schmitt_2012} for most distribution-extraction problems \cite{HOCKER1996469}. With sufficient data coverage and sufficiently flexible functional representations, they can recover substantial information about the underlying distribution. We do not seek to replace these methods, but instead address a complementary question that arises before inversion: with what degree of locality can a given hard-scattering channel and observable transfer information about the underlying distribution to the experimental level? If this locality can be calculated in advance \cite{Wang_2018,BALL2011112}, different processes and observables can be compared before analyzing the data, allowing measurement strategies that produce weaker smoothing of potential narrow peaks, dips, or rapid variations to be prioritized. Such fine structures may often signal new effects and physics. 

This question was motivated by our previous study of background photon distributions at the EIC \cite{li2026}. For the final-state photon-radiation observable constructed there, the experimentally measured variables completely determine the coordinate entering the background photon distribution, allowing the background distribution to be extracted point by point through the ratio of the experimental cross section to the complete theoretical baseline. This result demonstrates that certain collider observables can realize a local mapping from measurement space to distribution space. That specific process, however, does not establish the general factorization conditions required for such exact locality, nor does it provide a quantitative resolution scale when a measurement point corresponds to a finite region of the distribution. 

To this end, we define the theory baseline and the corresponding normalized direct-extraction quantity 
\begin{equation} 
F_{\rm dir}(X) = \frac{1}{B(X)} \frac{d\sigma}{dX}, \label{eq:direct_extraction} 
\end{equation} 
where $B(X) = \int d\Phi\, H(X,\Phi)\neq 0.$ We show that $F_{\rm dir}(X)$ is generally equal to a weighted average of the target distribution with respect to a sampling kernel jointly induced by the hard-scattering weight and the kinematic mapping. The width of this kernel in distribution space, or its covariance matrix in multiple dimensions, defines a locality diagnostic of the process–observable combination within a specified factorization setup. When $\xi(X,\Phi)=\xi_X$, the sampling kernel collapses to a point and exact direct extraction is realized; for a kernel of finite width, its covariance together with the local curvature of the target distribution controls the leading extraction bias. The framework therefore converts the question of whether “one measurement point corresponds to one point or to a finite region” into a calculable and comparable physical quantity. Candidate process--observable combinations can then be ranked before fitting: for the same target coordinate and physical feature scale, a smaller sampling-kernel width implies weaker forward smearing and hence better preservation of local structure. The previously constructed EIC photon-radiation observable corresponds to the zero-width limit of this framework.

\section{Direct Extractability and Distribution-Space Sampling}

After selecting \(F\) as the target distribution, \(H(X,\Phi)\) contains all calculable or externally constrained inputs, including hard scattering, source terms, propagators, phase space, Jacobians, and any other distributions treated as known inputs. Hence $B(X)$ is the complete theoretical baseline, not necessarily a lowest-order partonic cross section. Exact direct extraction corresponds to a special mapping structure: within the physical support of $H(X,\Phi)$, once the measured variables $X$ are fixed, the coordinate entering the target distribution is independent of all unobserved degrees of freedom. Thus, for all relevant unobserved configurations satisfying $H(X,\Phi)\neq 0$,
\begin{equation}
\xi(X,\Phi)=\xi_X .
\label{eq:exact_locality_condition}
\end{equation}
All relevant unobserved configurations therefore probe the same point of $F$, although they carry different theoretical weights. Using the definitions of $B(X)$ and $F_{\rm dir}(X)$ introduced in the Introduction, one immediately obtains
\begin{equation}
\frac{d\sigma}{dX}
=
F(\xi_X)
\int d\Phi\,H(X,\Phi)
=
B(X)F(\xi_X),
\label{eq:exact_factorization}
\end{equation}
and hence
\begin{equation}
F_{\rm dir}(X)=F(\xi_X).
\label{eq:exact_direct_extraction}
\end{equation}
When Eq.~\eqref{eq:exact_locality_condition} holds, \(F\) factors out of the \(\Phi\) integral and \(F_{\rm dir}(X)=F(\xi_X)\). Exact locality is a property of the kinematic map \(\xi(X,\Phi)\); any residual \(\Phi\) dependence produces smearing in distribution space.

In the following, we restrict the probabilistic sampling interpretation to factorized representations for which $H(X,\Phi)\geq 0$ over its physical support. Under this condition, the normalized weight
\begin{equation}
p_X(\Phi)
\equiv
\frac{H(X,\Phi)}{B(X)}
\label{eq:normalized_sampling_weight}
\end{equation}
defines a probability density, with $\int d\Phi p_X(\Phi)=1$. Some fixed-order or subtraction-based representations contain signed local weights. The exact-locality condition remains applicable, while the probabilistic interpretation of finite-width locality adopted below is restricted to nonnegative responses. To make explicit the distribution-space response induced by the physical hard process and kinematic mapping, we construct the normalized response kernel \cite{10.1111/j.1365-246X.1968.tb00216.x}
\begin{equation}
\rho_X(\zeta)
\equiv
\int d\Phi\,
p_X(\Phi)\,
\delta\!\left[\zeta-\xi(X,\Phi)\right].
\label{eq:distribution_sampling_kernel}
\end{equation}
By construction, $\int d\zeta\,\rho_X(\zeta)=1,$ and the quantity obtained after division by the complete theoretical baseline can be written as
\begin{equation}
F_{\rm dir}(X)
\equiv
\frac{1}{B(X)}
\frac{d\sigma}{dX}
=
\int d\zeta\,
\rho_X(\zeta)\,
F(\zeta).
\label{eq:direct_extraction_kernel_average}
\end{equation}
A collider observable can therefore be understood as a response-weighted sampling operation on the underlying distribution. For a general observable, $\rho_X(\zeta)$ has a finite width, and a fixed experimental measurement point effectively averages information from a finite region of the target distribution. By contrast, in the exact direct-extraction limit represented by Eq.~\eqref{eq:exact_locality_condition},
\begin{equation}
\rho_X(\zeta)
=
\delta(\zeta-\xi_X),
\label{eq:delta_kernel_limit}
\end{equation}
which immediately reproduces Eq.~\eqref{eq:exact_direct_extraction}. For a nonnegative sampling weight, this formulation also provides a stronger interpretation of Eq.~\eqref{eq:exact_locality_condition}: if the sampling kernel degenerates into a delta function, then $\xi(X,\Phi)$ is equal to the same value $\xi_X$ almost everywhere on the physical support of $H$. Exact locality is therefore a property of the complete factorized measurement mapping rather than of any particular parametrization of the target distribution $F$.

More importantly, for a specified factorization scheme, scale choice, and perturbative order, \(\rho_X\) is determined entirely by the observable, the kinematic mapping \(\xi(X,\Phi)\), and the calculable theoretical weight, without requiring prior knowledge of the unknown distribution \(F\). This allows the theoretical resolution scale provided by the observable and the hard process themselves to be clearly separated from the physical variation scale intrinsic to the target distribution.

Direct extractability is therefore determined by whether the distribution-space region mapped from fixed $X$ is finite or collapses to a point, equivalently whether $\rho_X$ has finite width or becomes a delta function. For finite kernels, their width and higher moments quantify nonlocal information mixing.

\section{Locality and Controlled Direct Extraction}

For a general collider observable, the distribution-space sampling kernel $\rho_X$ typically has a finite width. Consequently, even after dividing the experimental cross section by the complete theoretical baseline $B(X)$, the resulting quantity is no longer exactly the value of $F$ at a single definite coordinate, but rather a local weighted average in distribution space.

To quantify this departure from exact locality, we first define the center of the sampling kernel,
\begin{equation}
\bar{\xi}_X
\equiv
\int d\zeta\,
\rho_X(\zeta)\,
\zeta,
\label{eq:sampling_kernel_center}
\end{equation}
and the corresponding local variance,
\begin{equation}
\Delta\xi_X^2
\equiv
\int d\zeta\,
\rho_X(\zeta)
\left(\zeta-\bar{\xi}_X\right)^2.
\label{eq:locality_width}
\end{equation}
Here, $\Delta\xi_X$ is neither the statistical uncertainty of the experimental cross section nor the detector resolution. It is the distribution-space locality width jointly determined by the observable itself and the hard part of the corresponding scattering process. It characterizes the spread of the distribution-space region sampled by a fixed experimental measurement point within the specified factorization setup. When $\Delta\xi_X=0,$ exact direct extraction is recovered, whereas $\Delta\xi_X>0$ but sufficiently small corresponds to an approximately local observable. Let $\zeta=\bar{\xi}_X+\delta\xi,$ and expand the target distribution around the weighted center $\bar{\xi}_X$:
\begin{equation}
F(\zeta)
=
F(\bar{\xi}_X)
+
\delta\xi\,F'(\bar{\xi}_X)
+
\frac{\delta\xi^2}{2}
F''(\bar{\xi}_X)
+\cdots .
\label{eq:local_taylor_expansion}
\end{equation}
Using Eq.~\eqref{eq:direct_extraction_kernel_average}, the linear term vanishes because \(\langle\delta\xi\rangle_X=0\), giving
\begin{equation}
F_{\rm dir}(X)
=
F(\bar{\xi}_X)
+
\frac{\Delta\xi_X^2}{2}
F''(\bar{\xi}_X)
+\cdots .
\label{eq:leading_locality_correction}
\end{equation}
The leading local correction to exact direct extraction is thus controlled by two independent factors: the locality width $\Delta\xi_X$ in distribution space determined by the observable and its hard part, and the local curvature $F''(\bar{\xi}_X)$ of the target distribution within the sampled region. This is also why the weighted mean is the natural choice of local coordinate: if an arbitrary reference point were used instead, an extraction bias linear in the local spread would generally remain.

The expansion above can further be promoted to an error bound that does not rely on the asymptotic form of the expansion. Let $I_X$ denote an interval containing the relevant sampling region, and assume that the target distribution $F$ is twice differentiable on this interval. Taylor's theorem with remainder gives
\begin{equation}
\left|
F_{\rm dir}(X)-F(\bar{\xi}_X)
\right|
\leq
\frac{\Delta\xi_X^2}{2}
\sup_{\xi\in I_X}
\left|
F''(\xi)
\right|.
\label{eq:locality_error_bound}
\end{equation}
Thus, ``approximately local'' is no longer merely a qualitative description. Once the locality width induced by the observable is known, the direct-extraction error can be rigorously controlled for a class of distributions with a given upper bound on the local curvature. Importantly, calculating $\Delta\xi_X$ does not require prior knowledge of $F$: it depends only on the experimentally measured variables, the kinematic mapping $\xi(X,\Phi)$, and the calculable theoretical weight $H(X,\Phi)$. The intrinsic variation scale of the unknown distribution determines only how fine a physical structure can ultimately be resolved by a given observable. For example, for a local structure of characteristic width $w$ and characteristic amplitude comparable to the local value of $F$, one may estimate parametrically
\begin{equation}
\frac{|F''|}{|F|}
\sim
\frac{1}{w^2}.
\label{eq:curvature_structure_scale}
\end{equation}
Under this parametric estimate, the relative bias generated by distribution-space nonlocality scales as
\begin{equation}
\frac{|F_{\rm dir}(X)-F(\bar\xi_X)|}{|F(\bar\xi_X)|}
\sim
\frac{1}{2}
\left(
\frac{\Delta\xi_X}{w}
\right)^2.
\label{eq:relative_locality_bias}
\end{equation}
This yields the direct resolution condition $\Delta\xi_X\ll w.$ This condition does not imply that the experiment has infinite resolution. Real measurements remain affected by finite binning, statistical uncertainties, detector smearing, and event reconstruction. Rather, it states that when the distribution-space width induced by the observable itself is much smaller than the characteristic width of the target structure, integration over unobserved theoretical degrees of freedom does not introduce an additional smoothing effect of the same order.

It should be noted that the numerical value of $\Delta\xi_X$ itself changes under a reparametrization of the target-distribution coordinate, just as an ordinary variance does. Exact locality, however, does not suffer from this issue: under any smooth one-to-one change of variables, a delta-function sampling kernel remains a delta function. For approximately local observables, under a local change of variables, both the sampling width and the characteristic width of the target structure are rescaled by the same Jacobian. Therefore, at leading order, $\frac{\Delta\xi_X}{w}$ remains invariant. The quantity that truly characterizes local resolving power is therefore not the isolated value of $\Delta\xi_X$, but its dimensionless ratio to the physical variation scale of the target distribution.

For the multidimensional distributions more commonly encountered in practice, the above structure generalizes directly. Let the target-distribution coordinates be $\boldsymbol{\xi}=(\xi_1,\xi_2,\ldots),$ and define the weighted center $\bar{\xi}_{a,X}=\left\langle\xi_a\right\rangle_X$ and the covariance matrix
\begin{equation}
\Sigma_{ab}(X)
\equiv
\left\langle
\left(\xi_a-\bar{\xi}_{a,X}\right)
\left(\xi_b-\bar{\xi}_{b,X}\right)
\right\rangle_X.
\label{eq:locality_covariance_matrix}
\end{equation}
The covariance matrix $\Sigma\equiv[\Sigma_{ab}]$ is the multidimensional generalization of the one-dimensional locality variance $\Delta\xi_X^2$. For suitably scaled coordinates, the projected locality variance along a physically relevant direction $v$ is
\begin{equation}
\Delta_v^2(X)\equiv v^T\Sigma(X)v .
\label{1}
\end{equation}
For strongly non-Gaussian, multimodal, or long-tailed kernels, the full sampling kernel contains locality information beyond that captured by the covariance alone. Expanding around the weighted center gives
\begin{equation}
F_{\rm dir}(X)
=
F(\bar{\boldsymbol{\xi}}_X)
+
\frac{1}{2}
\Sigma_{ab}(X)\,
\partial_a\partial_b
F(\bar{\boldsymbol{\xi}}_X)
+\cdots ,
\label{eq:multidimensional_locality_expansion}
\end{equation}
where repeated indices are summed. Thus, the covariance matrix \(\Sigma_{ab}\) is the natural multidimensional generalization of the one-dimensional locality variance \(\Delta\xi_X^2\).

In a local linear approximation around the weighted center $\bar{\Phi}_{\alpha,X} \equiv \langle\Phi_\alpha\rangle_X, $ define $ \delta\Phi_\alpha \equiv \Phi_\alpha-\bar{\Phi}_{\alpha,X}$ and $ \delta\xi_a \equiv \xi_a-\bar{\xi}_{a,X}, $ and evaluate
\begin{equation}
    J_{a\alpha}(X) \equiv \left. \frac{\partial\xi_a(X,\Phi)} {\partial\Phi_\alpha} \right|_{\Phi=\bar{\Phi}_X}.
\end{equation}
Then $ \delta\xi_a \simeq J_{a\alpha}(X)\delta\Phi_\alpha. $ With $(C_\Phi)_{\alpha\beta} = \langle \delta\Phi_\alpha\delta\Phi_\beta \rangle_X, $ the locality covariance becomes
\begin{equation}
\Sigma
\simeq
J\,C_\Phi\,J^{T}.
\label{eq:covariance_propagation}
\end{equation}
It separates the local smearing into the kinematic sensitivity \(J\) of the target-distribution coordinates to unobserved variables and their covariance \(C_\Phi\).

\section{Observable Design and Exact Realization}

We now use the response kernel to design and compare process--observable combinations \cite{Arratia_2022}. Rather than starting from a fixed choice of measured variables, we ask which hard-scattering channel and observable definition minimize the dependence of $\boldsymbol{\xi}(X,\Phi)$ on the unobserved kinematics.

For a chosen target distribution $F(\boldsymbol{\xi})$, the practical procedure is as follows. One first enumerates candidate hard processes and observable definitions, specifying the measured variables $X$, the unobserved or integrated degrees of freedom $\Phi$, and the relevant cuts and binning. For each candidate, one then determines the kinematic map $\boldsymbol{\xi}=\boldsymbol{\xi}(X,\Phi)$ and constructs the corresponding theoretical response $H(X,\Phi)$.

For a given candidate, exact locality requires $\boldsymbol{\xi}(X,\Phi)$ to be independent of all unmeasured variables over the contributing phase space. A practical local test is
\begin{equation}
\frac{\partial \xi_a}{\partial \Phi_\alpha}=0,
\end{equation}
with discrete channels checked separately. Otherwise, the residual nonlocality is quantified by the response kernel $\rho_X$ or its locality covariance $\boldsymbol{\Sigma}$.

In the approximately local case, if $\boldsymbol{\xi}(X,\Phi)$ can be linearized over the relevant region, the locality covariance is approximately given by Eq.~\eqref{eq:covariance_propagation}. Locality can therefore be improved by reducing either contribution: one may choose kinematic variables that make $J$ smaller, or restrict the spread of the relevant unobserved degrees of freedom through appropriate event selection. When the linear approximation is inadequate, the full response kernel should be used.

Candidate process--observable combinations can then be ranked at the same target coordinate and physical feature scale: a smaller projected locality width, or a smaller $\Delta\xi_X/w$ in one dimension, implies weaker forward smearing and better preservation of local structure.

This allows $\Sigma_X$, or $\Delta\xi_X^2$ in one dimension, to serve as a new observable-quality measure \cite{PhysRevD.95.073002}. The design of collider observables traditionally requires balancing cross-section size, statistical sensitivity \cite{Diehl1994}, background contamination, detector acceptance, reconstruction resolution, and perturbative theoretical stability \cite{Arratia_2022}. For distribution extraction, an additional independent criterion can be introduced: ``distribution-space locality.'' For suitably scaled coordinates, Eq.~\eqref{1} provides a second-moment measure of locality along a physically relevant direction $v$.

To make the consequence of distribution-space locality for observable selection quantitatively explicit, we now adopt a forward-mapping perspective in a controlled one-dimensional Gaussian benchmark: a localized structure is first specified in the underlying distribution and then propagated through a Gaussian sampling kernel into the baseline-normalized quantity \(F_{\rm dir}\). Because this convolution is analytically tractable, the benchmark shows directly how the sampling-kernel variance controls the broadening and peak-amplitude suppression of the original structure.
 Consider an observable that induces a Gaussian response kernel in distribution space,
\begin{equation}
\rho_X(\zeta)
=
\frac{1}{\sqrt{2\pi}\,\Delta_X}
\exp\left[
-\frac{(\zeta-\bar{\xi}_X)^2}{2\Delta_X^2}
\right],
\end{equation}
where \(\Delta_X\) denotes its locality width. Let the target distribution contain a localized structure of intrinsic width \(w\) and amplitude \(A\) on top of a locally constant background \(F_0\),
\begin{equation}
F(\zeta)
=
F_0+
A\exp\left[
-\frac{(\zeta-\xi_0)^2}{2w^2}
\right].
\end{equation}
Using Eq.~(\ref{eq:direct_extraction_kernel_average}), the convolution can be evaluated analytically,
\begin{equation}
F_{\rm dir}(X)
=
F_0+
A\frac{w}{\sqrt{w^2+\Delta_X^2}}
\exp\left[
-\frac{(\bar{\xi}_X-\xi_0)^2}
{2(w^2+\Delta_X^2)}
\right].
\end{equation}
The response therefore broadens the structure to
\begin{equation}
w_{\rm eff}
=
\sqrt{w^2+\Delta_X^2},
\end{equation}
while reducing its peak amplitude to
\begin{equation}
A_{\rm eff}
=
A\frac{w}{\sqrt{w^2+\Delta_X^2}}.
\end{equation}
For \(\Delta_X\ll w\), the localized structure is essentially preserved, whereas for \(\Delta_X\gtrsim w\) it is substantially broadened and suppressed. Hence, if two observables \(X_A\) and \(X_B\) probe the same target distribution but satisfy $\Delta_A\ll w$ and $\Delta_B\gtrsim w,$ then \(X_A\) retains the narrow structure while \(X_B\) significantly smooths it in the theoretical forward mapping. Under otherwise comparable experimental conditions, the locality criterion therefore favors \(X_A\) for resolving structures on the scale \(w\).

This general structure has a particularly simple exact realization in the EIC photon-radiation observable studied previously \cite{li2026}. For that process, the experimentally retained variables can be chosen as $X=(y_\gamma,k_\perp),$ where $y_\gamma$ and $k_\perp=|\textbf{k}_\perp|$ are the rapidity and transverse momentum of the final-state photon, respectively. The background-photon enhancement factor for the radiation is
\begin{equation}
F(X)=
1+n(y_\gamma,k_\perp).
\label{eq:eic_enhancement_factor}
\end{equation}
It originates from Bose enhancement of photon radiation by an electron in a photon field. The term $1$ represents spontaneous radiation, whereas $n$ represents stimulated radiation. All remaining unobserved kinematic variables are collectively included in $\Phi$. The factorization relation established in our previous EIC analysis \cite{li2026} can be summarized as
\begin{equation}
\frac{d\sigma_{\rm EIC}}{dX}
=
\left[
\int d\Phi\,
H_{\rm EIC}(X,\Phi)
\right]
\left[1+n(X)\right],
\label{eq:eic_exact_factorization}
\end{equation}
where $1+n(X)$ is independent of all unobserved variables $\Phi$. The coordinate entering the target distribution is therefore completely fixed by the measured variables $X$: $\boldsymbol{\xi}(X,\Phi)=X.$ In the language of distribution-space sampling, this means
\begin{equation}
\rho_X(\boldsymbol{\xi})
=
\delta^{(2)}
\!\left(
\boldsymbol{\xi}-X
\right),
\label{eq:eic_delta_sampling_kernel}
\end{equation}
where \(\delta^{(2)}\) reflects the two-dimensional target coordinate; correspondingly, \(d\zeta\) is understood as \(d^2\zeta\) in this case, and hence $\Sigma_{\rm EIC}=0.$ The previously obtained relation
\begin{equation}
F_{\rm dir}(X)
=
\frac{d\sigma_{\rm EIC}/dX}
     {B_{\rm EIC}(X)}
=
1+n(X)
\label{eq:eic_effective_distribution}
\end{equation}
is the zero-width limit of Eq.~\eqref{eq:direct_extraction_kernel_average}, with \(\rho_X=\delta^{(2)}(\boldsymbol{\xi}-X)\) and \(\boldsymbol{\Sigma}_{\rm EIC}=0\) at truth level within the stated signal model and approximations.

\section{Summary and Outlook}

We have introduced a distribution-space response kernel for factorized observables of the form in Eq.~\eqref{eq:factorized_observable}, with the aim of diagnosing how locally a measured point \(X\) probes a selected target distribution before any global inversion is performed. At fixed \(X\), the kernel specifies the distribution coordinates \(\boldsymbol{\xi}\) entering the theoretical response and can be calculated without assuming the global shape of the unknown target distribution. A delta-function kernel corresponds to exact pointwise extraction, whereas a finite kernel produces a response-weighted average over distribution space. Its variance or covariance provides a useful second-moment characterization of this nonlocality, while the full kernel retains additional information for non-Gaussian, multimodal, or long-tailed responses.

The framework provides both a locality criterion and a practical workflow for process--observable design. After fixing the target distribution, candidate hard processes, measured variables, integrated degrees of freedom, cuts, and binning can be mapped to their corresponding distribution-space kernels and compared before fitting or unfolding. For the same target coordinate and physical feature scale, a smaller locality width implies weaker forward smearing and better preservation of local structure. Distribution-space locality therefore supplies an additional observable-quality criterion, complementary to statistical sensitivity, detector resolution, background control, and theoretical stability. Within the stated signal model and approximations, the EIC photon-radiation example realizes the zero-width limit at truth level, while the finite-width DIS realization explicitly illustrates how increasing kernel width suppresses and broadens a localized structure.

A natural next step is to extend the present finite-width DIS realization to processes in which the distribution-space width arises from genuinely unobserved multidimensional kinematics and competing hard-scattering channels.

%\section*{Acknowledgments}
\appendix
\section{Inclusive DIS: Observable Choice and Forward Transmission of Local Features}
\label{sec:endmatter_forward_resolution}

As a concrete realization of the general construction in the main text, consider unpolarized inclusive DIS at fixed $Q^2$. In one-photon exchange and at leading twist and lowest order in QCD with massless quarks,
\begin{equation}
\frac{d^2\sigma}{dx\,dQ^2}
=
H_{\rm DIS}(x,Q^2)\,F(x,Q^2),
\label{eq:dis_factorized}
\end{equation}
where $F(x,Q^2)\equiv\frac{F_2(x,Q^2)}{x}.$ denotes the target distribution entering the locality analysis and 
\begin{equation}
H_{\rm DIS}(x,Q^2)
=
\frac{2\pi\alpha^2}{Q^4}
\left[
2-\frac{2a}{x}+\frac{a^2}{x^2}
\right],
\qquad
a=\frac{Q^2}{s}.
\label{eq:dis_hard}
\end{equation}
Here $x$ is the Bjorken scaling variable, $Q^2$ is the photon virtuality, and $s$ is the squared center-of-mass energy of the incoming lepton--proton system.Thus Eq.~\eqref{eq:dis_factorized} has precisely the form of the general response $H\,F[\xi]$ used in the main text, with the target coordinate $\xi=x$.

Now integrate the same cross section over a finite interval $B=[\ell,u]$,
\begin{equation}
\frac{d\sigma_B}{dQ^2}
=
\int_\ell^u dx\,
H_{\rm DIS}(x,Q^2)F(x,Q^2).
\label{eq:dis_integrated}
\end{equation}
Following the normalization introduced in the main text,
\begin{equation}
\mathcal N_B(Q^2)
=
\int_\ell^u dx\,H_{\rm DIS}(x,Q^2),
\label{eq:dis_normalized}
\end{equation}
Equation~\eqref{eq:dis_normalized} can therefore be written directly in the sampling-kernel form
\begin{equation}
F_{{\rm dir},B}(Q^2)
=\frac{d\sigma_B/dQ^2}{\mathcal N_B(Q^2)}=
\int dx\,\rho_B(x)F(x,Q^2),.
\label{eq:dis_sampling_kernel}
\end{equation}
where $\rho_B(x)=\frac{H_{\rm DIS}(x,Q^2)}{\mathcal N_B(Q^2)}.$ This is the explicit DIS realization of the distribution-space kernel defined in the main text. It is positive and normalized and is fixed completely by the calculable hard response and the observable interval.

Because the overall factor $2\pi\alpha^2/Q^4$ cancels from the normalized kernel, its moments can be obtained analytically. Defining
\begin{equation}
n_B
=
\left[
2x-2a\ln x-\frac{a^2}{x}
\right]_\ell^u,
\label{eq:dis_nB}
\end{equation}
one finds
\begin{equation}
\bar x_B
=
\frac{
\left[
x^2-2ax+a^2\ln x
\right]_\ell^u
}{n_B},
\label{eq:dis_mean}
\end{equation}
and
\begin{equation}
(\Delta x_B)^2
=
\frac{
\left[
\frac{2}{3}x^3-ax^2+a^2x
\right]_\ell^u
}{n_B}
-
\bar x_B^2.
\label{eq:dis_variance}
\end{equation}
These are exactly the kernel mean and locality width introduced in the main text.

To examine how this locality width affects a localized structure in the forward direction, take at fixed $Q^2=Q_*^2$
\begin{equation}
F_{\rm test}(x,Q_*^2)
=
F_{\rm ref}(x,Q_*^2)
+
A\exp\left[
-\frac{(x-x_0)^2}{2w^2}
\right].
\label{eq:dis_test_feature}
\end{equation}
The Gaussian term is only a localized probe of the response. Substituting Eq.~\eqref{eq:dis_test_feature} into the forward map gives
\begin{equation}
F_{{\rm dir},B}[F_{\rm test}]=F_{{\rm dir},B}[F_{\rm ref}]+A\eta_B(x_0,w).
\label{eq:dis_forward_feature}
\end{equation}
The input amplitude $A$ is therefore reduced by the observable averaging. We quantify the retained fraction by
\begin{equation}
\eta_B(x_0,w)
\equiv
\int dx\,\rho_B(x)
\exp\left[
-\frac{(x-x_0)^2}{2w^2}
\right],
\label{eq:dis_retention}
\end{equation}
so that the transmitted feature has amplitude $A\eta_B(x_0,w)$. Choosing $x_0=\bar x_B$ gives the centered retention factor $\eta_B(w)$. In direct correspondence with the resolution scale introduced in the main text, define
\begin{equation}
w_p(B)
=
\inf\left\{w>0:\eta_B(w)\ge p\right\}.
\label{eq:dis_wp}
\end{equation}

\begin{figure}[t]
\includegraphics[width=0.5\columnwidth]{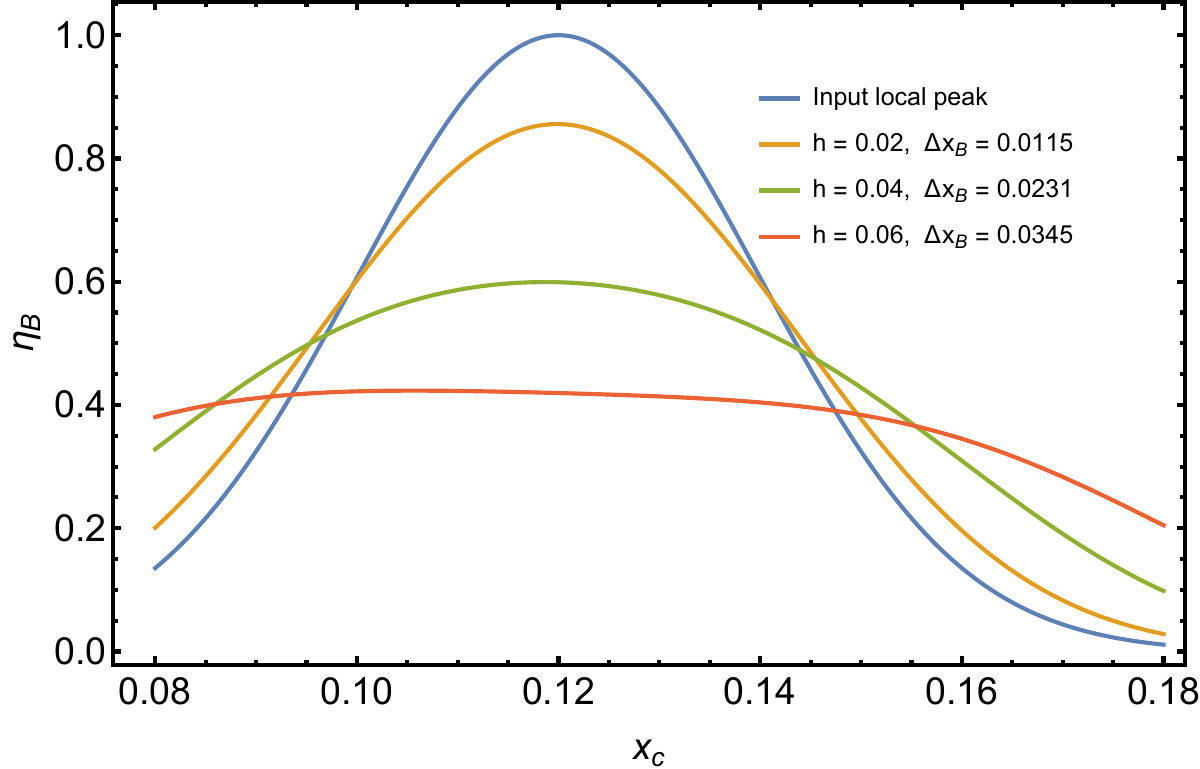}
\caption{Forward smearing of the same localized Gaussian feature in inclusive DIS. The input is centered at \(x_0=0.12\) with width \(w_0=0.020\). The three forward responses correspond to the intervals \(B_1=[0.10,0.14]\), \(B_2=[0.08,0.16]\), and \(B_3=[0.06,0.18]\), with kernel widths \(\Delta x_B=0.01154\), \(0.02305\), and \(0.03448\), respectively. Larger kernel variance produces a lower and broader transmitted peak.}
\label{fig:dis_forward_smeared_peaks}
\end{figure}

For a representative EIC realization, we take a $5~\mathrm{GeV}$ electron beam colliding with a $41~\mathrm{GeV}$ proton beam, corresponding to $\sqrt{s}\simeq 28.6~\mathrm{GeV}$, and fix $Q_*^2=10~\mathrm{GeV}^2.$ To vary the degree of observable coarse graining in a controlled manner, we keep the bin center fixed at $x_c=0.12$ and consider the one-parameter family $B_h=[x_c-h,x_c+h],$ with three equally spaced half-widths, $h=0.02, 0.04, 0.06,$ corresponding to $B_1=[0.10,0.14], B_2=[0.08,0.16], B_3=[0.06,0.18].$ The analytic kernel widths are $\Delta x_{B_1}=0.0115,\Delta x_{B_2}=0.0231, \Delta x_{B_3}=0.0345.$ Numerically evaluating the full forward response in Eq.~\eqref{eq:dis_retention} gives $w_{90}(B_1)=0.0246, w_{90}(B_2)=0.0491, w_{90}(B_3)=0.0735.$ The ratio remains nearly constant,
\begin{equation}
\frac{w_{90}}{\Delta x_B}\simeq2.13,
\label{eq:dis_scaling}
\end{equation}
throughout this controlled bin-width scan. Thus increasing the coarse-graining scale increases the distribution-space kernel width and correspondingly increases the minimum feature width required for a fixed level of forward retention. Equivalently, for a localized structure of fixed intrinsic width $w$, the larger-variance kernel produces stronger attenuation, whereas reducing the bin width continuously approaches the local differential limit. Figure~\ref{fig:dis_forward_smeared_peaks} provides a direct forward visualization of locality loss: for the same underlying Gaussian feature, increasing the DIS kernel width lowers the transmitted peak and broadens it, yielding a flatter smeared image.

%Figure~\ref{fig:dis_forward_smeared_peaks} provides a direct forward visualization of locality loss: for the same underlying Gaussian feature, increasing the DIS kernel width lowers the transmitted peak and broadens it, yielding a flatter smeared image.

This example is restricted to the lowest-order DIS representation in Eq.~\eqref{eq:dis_factorized}. Beyond this approximation, coefficient-function convolutions, additional partonic channels, finite binning in other variables, QED radiation, and detector response generate further sources of nonlocality. Its role here is simply to provide a physical realization of the general sequence developed in the main text:
\begin{equation}
H
\longrightarrow
\rho
\longrightarrow
(\bar\xi,\Delta\xi)
\longrightarrow
\text{forward transmission of local structure}.
\end{equation}

\bibliography{ref}

\end{document}